\documentclass[a4paper,aps,pre,floatfix,twocolumn,nofootinbib,showpacs,superscriptaddress]{revtex4-1}

\usepackage{times}
\usepackage{verbatim}
\usepackage{bbold}
\usepackage{bbm}
\usepackage[pdftex]{graphicx}
\setkeys{Gin}{width=0.9\columnwidth}
\usepackage{latexsym,amsmath,verbatim}
\usepackage{color}
\usepackage{hyperref}
\usepackage{lipsum}
\usepackage{rotating}
\usepackage{multirow}
\usepackage[english]{babel}
\usepackage{microtype}
\usepackage{comment}
\usepackage[normalem]{ulem}

\usepackage{color}
\hypersetup{colorlinks=true,
  linkcolor=blue,
  urlcolor=blue,
  citecolor=blue,
  pdfhighlight=/N}

\newcommand{\av}[1]{\left\langle {#1} \right\rangle}

\newcommand{\ai}{\alpha_\mathrm{i}}
\newcommand{\aoo}{\alpha_\mathrm{o}}
\newcommand{\RSELF}{R_\infty(\mathrm{SELF})}
\newcommand{\ROTHER}{R_\infty(\mathrm{OTHER})}
\newcommand{\RNONE}{R_\infty(\mathrm{NONE})}
\newcommand{\RBOTH}{R_\infty(\mathrm{BOTH})}
\newcommand{\ImSELF}{I_m(\mathrm{SELF})}
\newcommand{\ImOTHER}{I_m(\mathrm{OTHER})}
\newcommand{\ImNONE}{I_m(\mathrm{NONE})}
\newcommand{\ImBOTH}{I_m(\mathrm{BOTH})}

\newcommand{\RIN}{R_\infty(\mathrm{SELF})}
\newcommand{\ROUT}{R_\infty(\mathrm{OTHER})}
\newcommand{\ImIN}{I_m(\mathrm{SELF})}
\newcommand{\ImOUT}{I_m(\mathrm{OTHER})}

\newcommand{\be}{\begin{equation}}
\newcommand{\ee}{\end{equation}}

\begin{document}

\title{The advantage of self-protecting interventions in mitigating
  epidemic circulation at the community level}

\author{Romualdo Pastor-Satorras}
\affiliation{Departament de F\'isica, Universitat Polit\`ecnica de
  Catalunya, Campus Nord B4, 08034 Barcelona, Spain}

\author{Claudio Castellano}

\affiliation{Istituto dei Sistemi Complessi (ISC-CNR), Via dei Taurini 19,
  I-00185 Rome, Italy}

\affiliation{Centro Ricerche Enrico Fermi, Piazza del Viminale, 1, I-00184
  Rome, Italy}

\begin{abstract}
  Protecting interventions of many types (both pharmaceutical and
  non-pharmaceutical) can be deployed against the spreading of a
  communicable disease, as the worldwide COVID-19 pandemic has
  dramatically shown.  Here we investigate in detail the effects at the
  population level of interventions that provide an asymmetric
  protection between the people involved in a single interaction.  Masks
  of different filtration types, either protecting mainly the wearer or
  the contacts of the wearer, are a prominent example of these
  interventions. By means of analytical calculations and extensive
  simulations of simple epidemic models on networks, we show that
  interventions protecting more efficiently the adopter (e.g the mask
  wearer) are more effective than interventions protecting primarily the
  contacts of the adopter in reducing the prevalence of the disease and
  the number of concurrently infected individuals (``flattening the
  curve'').  This observation is backed up by the study of a more
  realistic epidemic model on an empirical network representing the
  patterns of contacts in the city of Portland.  Our results point out
  that promoting wearer-protecting face masks and other self-protecting
  interventions, though deemed selfish and inefficient, can actually be
  a better strategy to efficiently curtail pandemic spreading.
\end{abstract}

\maketitle

\section{Introduction}

The recent COVID-19 pandemic has shown how fragile world societies are when
confronted to the runaway spreading of a new virus
strain~\cite{christakis2020apollo}.  In stark contrast with other diseases,
however, the combination of recent scientific advances, liberal funding by
governments and the availability of large scale manufacturing, has allowed to
create and deploy world-wide vaccines for the SARS-CoV-2 virus at unprecedented
speed~\cite{ball_vaccines_2021}, a fact that has without any doubts greatly
diminished the death toll expected in this pandemic. Despite this huge effort,
however, more than nine months lapsed between the declaration of the COVID-19
pandemic by WHO in March 2020 and the approval of the first vaccine in December
2020. The large-scale deployment of vaccines worldwide required additional
months and is still an unsolved problem in many areas of the planet.  Moreover,
the protection guaranteed by vaccines against recently emerged new variants is
far from perfect.  In this context, societies have had and still have to fight
the propagation of the virus by adopting also more traditional
non-pharmaceutical interventions~\cite{Perra.2021}.  These interventions, aiming
at limiting the spread of a disease~\cite{christakis2020apollo}, include
collective actions, such as school closures, prohibition of large meetings,
closing of public transportation, curfews or lockdown orders~\cite{npi_ecdc}. At
an individual level, non-pharmaceutical interventions include, among others,
physical distancing, hand and respiratory hygiene, and face masks.
Understanding the impact of each of these interventions at the level of the
single transmission event among individuals and how this translates at the
community level is a crucial goal, which has attracted a huge scientific
interest~\cite{Lin2020,Reguly2022,Gozzi2021,Jarvis2020,Moritz2020,Brauner2021,Davies2020,Stutt2020,Li2020,Althouse2020,Dehning2020,Hsiang2020,Yacong2021}.

In this paper we present an analysis of how the reduction of the chance that the
disease is transmitted in a single contact, induced by a generic type of
protecting intervention, results in an overall mitigation of the epidemic
circulation in the population.  In particular, we focus on asymmetric
interventions having different efficacies at the individual level, i.e., in
reducing the infectivity of an infected individual with respect to decreasing
the susceptibility of a susceptible one.

As one of the most widely adopted means to fight the current pandemic, mask
wearing is the paradigmatic example of these types of interventions.  Given the
evidence for the airborne transmission of SARS-CoV-2 by means of droplets and
aerosols~\cite{Greenhalgh:2021vb}, recent research has shown that face masks are
a very effective way to reduce the spread of
COVID-19~\cite{Mitze32293,Behring.2021,LIANG2020101751,EIKENBERRY2020293,NGONGHALA2020108364,Howarde2014564118,Watanabe2021,Robinson2022}.
Different types of masks exist, characterized by very different protecting
performance, both qualitatively and
quantitatively~\cite{Koh:2022up,Leith:2021uw}.  Some protect mainly the mask
wearer, such as valved masks endowed with an exhalation valve.  Others mainly
protect the contacts of the mask wearer, such as surgical or cloth masks. Such
asymmetric efficacy is however more general.
For example, vaccines reduce the susceptibility of individuals to being
infected upon contact with an infectious person, but they need not be
equally effective in reducing the capability of vaccinated infected
individuals to transmit the pathogen further if they ever catch the
disease~\cite{Braeye2021}.  Also, symptomatic treatment reducing cough
and sneezing hinders the possibility to transmit the disease onward but
it does not reduce the chance to be infected.

Here we present a comparative analysis of the global efficiency of
generic protective interventions (PI) in the context of epidemic models
on networks~\cite{PastorSatorras2015}, representing the patterns of social
contacts among people that mediate the propagation of an infective
pathogen~\cite{Jackson2010}. We use a formulation over networks with a
static topology, parametrizing the effect of PIs in terms of the
fraction of adopters and of the efficacy of adoption in reducing the
chance to receiving/transmitting the disease.

We analyze the effect of different types of PIs at the community level
by considering two paradigmatic models of disease propagation in
networks~\cite{Keeling07book,Kiss2017}: the Susceptible-Infected-Removed
(SIR) model for non-recurrent diseases, that confer immunity and develop
in outbreaks (similarly to COVID-19, at least on short time scales), and
the Susceptible-Infected-Susceptible (SIS) model for recurrent diseases
that may lead to a steady endemic state. Using a combination of
numerical simulations and analytical calculations we show that the
adoption of asymmetric PIs leads to surprisingly different scenarios. In
particular interventions protecting the adopter turn out to have a
stronger effect at the population level.  If PIs of the same efficacy in
reducing infectivity/susceptibility are available, it is better to use
PIs protecting the adopter, as they suppress more the global circulation
of the disease. We investigate the origin of this unexpected finding and
show that the conclusion holds also when considering more realistic
interaction patterns and disease dynamics.

\section{Results}
\subsection{Modeling the effect of protecting interventions}

The effect of protecting interventions on the propagation of a disease
can be modeled by considering the modification of the probability that
the disease is transmitted between an infected and a susceptible
individual~\cite{Miller2007, Behring.2021}. Consider a population of $N$
individuals, whose pattern of contacts is determined by a complex
network~\cite{Newman10}, in which nodes represent individuals and edges
the presence of social contacts between pairs of individuals.  The
network is fully described by its adjacency matrix $a_{ij}$, taking
value $1$ if nodes $i$ an $j$ are connected, and zero otherwise. From a
statistical perspective, the network can be characterized by its degree
distribution $P(k)$, representing the probability that a randomly chosen
node has degree $k$ (i.e., it is connected to $k$ other individuals).

Disease transmission is mediated by contacts between infected and susceptible
individuals. The properties of this transmission are mathematically
encoded~\cite{Keeling07book,Kiss2017} in terms of an infection rate $\beta$,
representing the probability per unit time that the infection is transmitted
from an infected to a susceptible along a single contact. Protecting
interventions induce a reduction of this infection rate in two possible
manners. An intervention involving the infected individual diminishes the
infection rate along any edge emanating from him/her by a factor $\aoo$. An
intervention affecting the susceptible individual decreases, by a factor $\ai$,
the infection rate of any edge pointing to him/her. For example, in the case of
airborne transmission, a mask may reduce the number of viral particles exhaled
by an infected individual and also reduce the number of viral particles that a
susceptible can inhale from the environment~\cite{Behring.2021}. As the example
of masks shows, the factors $\aoo$ and $\ai$ are not necessarily equal.  For
example valved masks strongly protect the wearer $\ai \ll 1$ while they do not
impede viral transmission towards the others $\aoo \approx 1$.  For cloth masks
the opposite is true.  We compare scenarios where a single type of PI (with
given $\aoo$ and $\ai$) is adopted by a fraction $f$ of the population.  For a
given edge pointing from node $i$ to node $j$, the infection rate associated to
the disease transmission from $i$ to $j$ takes the form
\begin{equation}
  \beta_{ij} = \beta a_{ij} \left[ 1 - m_i (1- \aoo) \right ]  \left[ 1 -
    m_j (1-  \ai)\right ],
  \label{eq:def_infection_rate_directed}
\end{equation}
where $m_i$ are quenched Bernoulli stochastic variables taking value $1$
with probability $f$ and zero with probability $1-f$.  From this
definition it is obvious that, even if the contact network is undirected
(symmetric $a_{ij}$)~\cite{Newman10}, the transmission network given
by~\eqref{eq:def_infection_rate_directed} is effectively directed, with
$\beta_{ij} \neq \beta_{ji}$ unless $f=0$ (no PI adoption in the
population) or $f=1$ (whole population adopting PI), which lead to
$\beta_{ij} = \beta a_{ij}$ and $\beta_{ij} = \beta \ai \aoo a_{ij}$,
respectively.  In these cases, the infection process is symmetric if the
underlying contact network is undirected.  In our study we are concerned
with the differences between PIs that mostly protect the adopter,
characterized by $\ai \ll \aoo$, and PIs that mostly protect his/her
contacts, given by $\aoo \ll \ai$. For the sake of simplicity, we focus
on what we call SELF interventions (with $\ai < 1$, $\aoo=1$) that offer
protection to the adopter and no protection for his/her contacts, and
OTHER interventions ($\ai = 1$, $\aoo<1$), that offer protection to the
contacts, and no protection to the adopter.

In the following, we investigate the effects of PIs in fundamental
models of disease propagation, either non-recurrent or recurrent.

\subsection{Non-recurrent diseases}
\label{sec:model}

As an example of a communicable non-recurrent disease (that is, a
disease that confers permanent immunity), we consider the simple
Susceptible-Infected-Removed (SIR) model~\cite{Keeling07book}. The SIR
model is defined in terms of three compartments. Susceptible individuals
are healthy and can contract the disease. Infected individuals carry the
disease and can infect susceptible ones upon contact with a rate
(probability per unit time) given by the factor $\beta_{ij}$ in
Eq.~(\ref{eq:def_infection_rate_directed}). In their turn, infected
individuals can recover and become removed with a constant rate
$\mu$. By rescaling time, we can absorb the rate $\mu$ and consider a
single parameter, the spreading rate $\lambda = \beta/ \mu$.

\begin{figure}[t]
  \centering
  \includegraphics[width=\columnwidth]{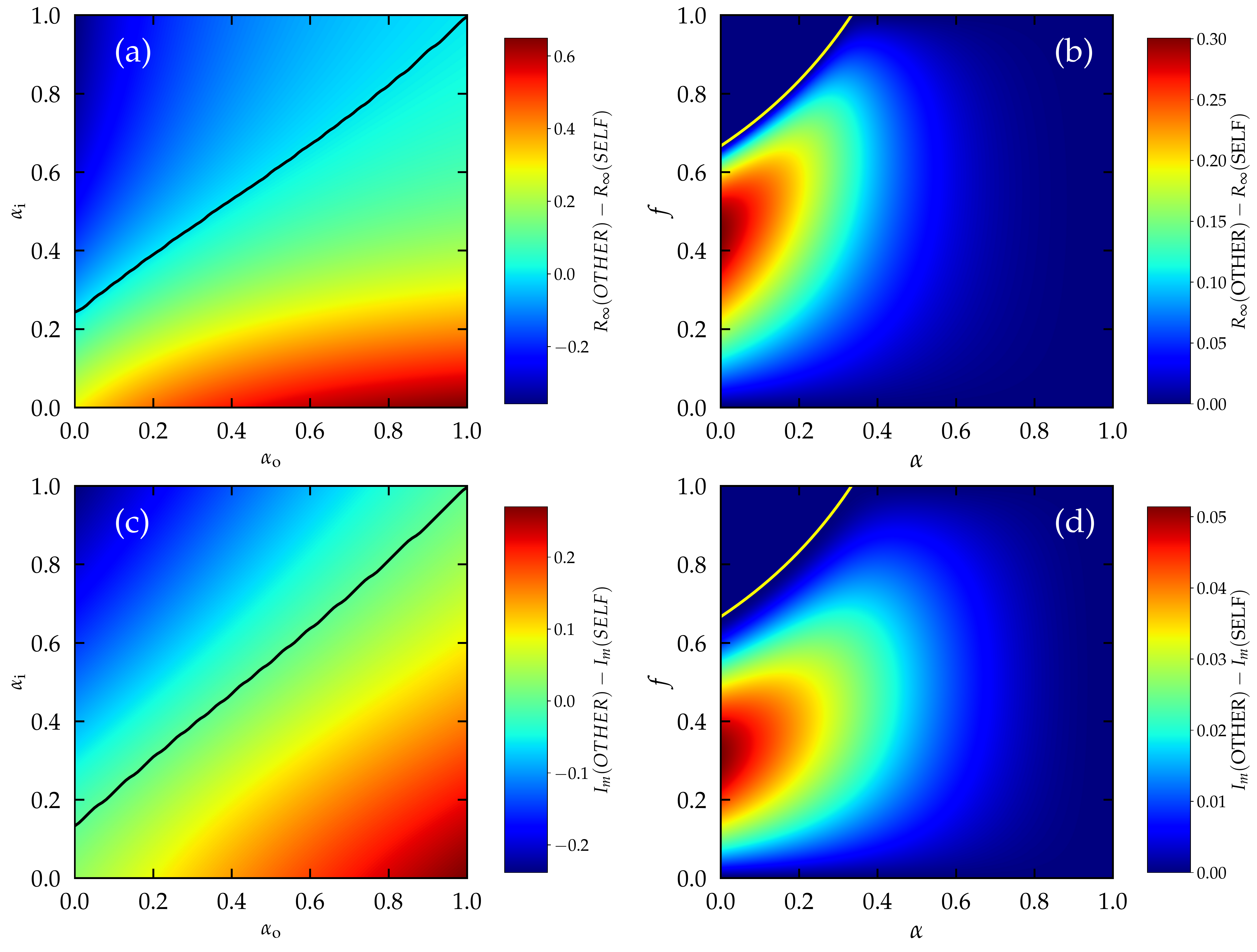}
  \caption{{\bf SELF interventions are more effective than OTHER interventions
      at the population level.}  (a) Difference between the final prevalence
    $R_\infty$ for the OTHER and the SELF scenario ($\ROTHER-\RSELF$), as a
    function of $\ai$ and $\aoo$ of the individual PI. Notice that $\ROTHER$
    depends only on $\aoo$, and $\RSELF$ depends only on $\ai$. The solid black
    line indicates the zero value.  (b) Difference between the final prevalence
    $R_\infty$ for the OTHER and the SELF scenario as a function of the fraction
    $f$ of adopters and the efficacy $\alpha$ of the individual PI.  The solid
    yellow line denotes where $\lambda_c(\alpha)=\lambda$.  Above it, the system
    is subcritical and $\ROTHER-\RSELF=0$.  (c) Difference between the maximum
    value of the incidence $I_{m}$ for the OTHER and the SELF scenario
    ($\ImOTHER-\ImSELF$), as a function of $\ai$ and $\aoo$ of the individual
    PI. Notice that $\ImOTHER$ depends only on $\aoo$, and $\ImSELF$ depends only
    on $\ai$. The solid black line indicates the zero value.  (d) Difference
    between the maximum value of the incidence $I_{m}$ for the OTHER and the
    SELF scenario as a function of the fraction $f$ of adopters and the efficacy
    $\alpha$ of the individual PI.  The solid yellow line denotes where
    $\lambda_c(\alpha)=\lambda$.  In all plots quantities greater than zero
    indicate that SELF interventions perform better. Solution of homogeneous MF
    equations for a network of degree $K=7$ and for $\lambda=0.5$.  In panels
    (a) and (c) $f=0.5$. For these parameters, in panels (a) and (c) both SELF
    and OTHER strategies operate above their respective thresholds.}
  \label{2dplot}
\end{figure}

Focusing for simplicity on homogeneous networks with a uniform degree
$k_i = K$, a mean-field theoretical analysis (see Methods~\ref{hmf_sir})
shows that the SIR model exhibits a transition between a phase with
non-extensive outbreaks only and a phase with macroscopic ones, at a
critical value of the spreading rate $\lambda$ (see
Methods~\ref{threshold})
\begin{equation}
  \label{eq:threshold_main}
  \lambda_c = \frac{1}{K-1} \frac{1}{ f \ai \aoo +1 - f}.
\end{equation}
From this expression, we see that the threshold is symmetric in the
parameters $\ai$ and $\aoo$, in agreement with the results in
Ref.~\cite{Behring.2021}, and that the maximum system-wide protection
is obtained, for a given value of adoption probability $f$, when
$\ai \aoo = 0$, which can be attained either in the perfect SELF scenario
($\ai=0$) or in the perfect OTHER case ($\aoo=0$).  In the same
theoretical framework, it is also possible to calculate the total
prevalence $R_\infty$, i.e., the overall fraction of individuals
infected by the disease throughout the outbreak, obtaining, slightly
above the threshold (see Methods~\ref{PrevalenceSIR})
\begin{equation}
  R_\infty \simeq  \frac{2 \Delta}{\lambda_c^2 (K-1)}\frac{f \ai +1 -
    f}{f \ai^2 \aoo + 1 - f},
  \label{eq:13}
\end{equation}
where $\Delta = \lambda-\lambda_c$ quantifies the distance from the critical
threshold.  This expression reveals that, at variance with the position of the
threshold, the size of an outbreak is not symmetric with respect to the efficacy
of PIs: At fixed values of $f$, $\lambda$ and of the threshold $\lambda_c$
(i.e.~for the product $\ai \aoo$ fixed to a constant $A$), if we substitute
$\aoo = A / \ai$, the total prevalence is easily shown to be an increasing
function of $\ai$, while setting $\ai = A / \aoo$ leads to a total prevalence
that decreases with $\aoo$.  In other words, while using more effective SELF
interventions (i.e., reducing $\ai$) decreases $R_\infty$, imposing more
effective OTHER interventions (i.e reducing $\aoo$) creates the opposite result,
an increase of $R_\infty$.  This signals that SELF interventions, protecting the
adopter, are more effective than OTHER interventions, protecting the contacts,
at reducing the overall spreading of the disease at the community level.  This
is confirmed by integrating numerically the homogeneous mean field
equations~\eqref{eveqin}-\eqref{eveqfin}, for generic values of $\ai$ and $\aoo$
(see Fig.~\ref{2dplot}(a)).
Most of the parameter space results in $\ROTHER>\RSELF$, i.e. a larger
circulation of the disease when OTHER interventions are adopted.

The comparison of SELF and OTHER scenarios provides remarkable insight
also at the level of the individual. As shown in Methods~\ref{relation},
the relation between the asymptotic prevalence $R_1(\infty)$ for adopting
individuals and the prevalence $R_0(\infty)$ for nonadopting ones is
\begin{eqnarray}
  R_1(\infty) = 1 - [1-R_0(\infty)]^{\ai},
  \label{R1R0}
\end{eqnarray}
which does not depend on the protection level $\aoo$ offered to contacts.
This means that adopting interventions offering more protection to
contacts (i.e. reducing $\aoo$) has exactly the same effect on the
prevalence of adopting and nonadopting individuals.  In particular, if
OTHER PIs are used, since $\ai=1$ then $R_1(\infty)=R_0(\infty)$, for
any value of $\aoo$.  In such a case, from the point of view of a given
individual, adopting an intervention is perfectly equivalent to not
adopting it, as the probability to be eventually infected is exactly the
same.  With the benefit of hindsight this result is not so odd: If
$\ai=1$, PIs do not provide any protection against the transmission from
others to the adopter, hence for the individual adopter the risk of
being infected cannot be smaller than for a non-adopter.  However,
Eq.~\eqref{R1R0} reveals that, while adopting a PI always implies some
degree of inconvenience (think for example at the bother of wearing a
mask), no immediate utility exists for the individual adopting an OTHER
PI, while such a utility exists for SELF PIs. See the concluding section
for further discussion.

\begin{figure}[t]
  \centering
  \includegraphics[width=\columnwidth]{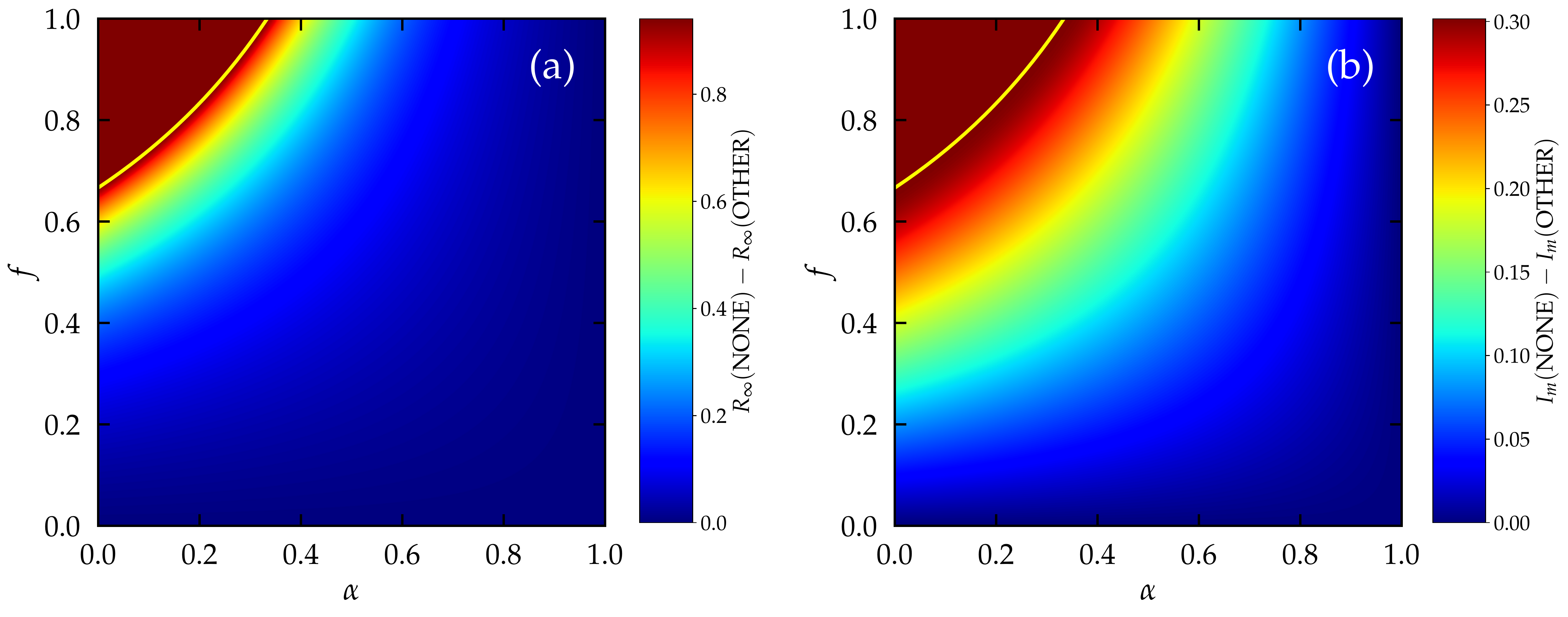}

  \caption{{\bf OTHER interventions flatten the curve but barely reduce prevalence
      with respect to no intervention.} (a) Difference between the final
    prevalence $R$ for the NONE and the OTHER scenario as a function of
    the fraction $f$ of adopters and the efficacy $\alpha$ of the
    individual PI.  (b) Difference between the maximum value of the
    incidence $I_{m}$ for the NONE and the OTHER scenario as a function of
    the fraction $f$ of adopters and the efficacy $\alpha$ of the
    individual PI. Solution of MF equations for a homogeneous network
    of degree $K=7$ for $\lambda=0.5$.  The solid yellow lines are where
    $\lambda_c(\alpha)=\lambda$ for the OTHER scenario.}
    \label{2dplot2}
\end{figure}

From now on we mostly focus on the comparison of the effects at the
population level of SELF and OTHER interventions with exactly the same efficacy
$\alpha$ ($\ai=\alpha$ for the SELF case, $\aoo=\alpha$ for the OTHER case).
For reference we consider also the case NONE, where no intervention is adopted,
corresponding to $\ai = \aoo = 1$, and the case BOTH, given by
$\ai=\aoo=\alpha$, and representing equally protecting performance in both
directions. In this last case (i.e., along the
diagonal in Fig.~\ref{2dplot}(a)) SELF PIs outperform OTHER PIs for any
value of the fraction $f$ of adopters.  This is checked by considering
$\ROTHER-\RSELF$ as a function of $\alpha$ and $f$.  The values plotted in
Fig.~\ref{2dplot}(b), which are nonnegative for any value of $\alpha$
and $f$, demonstrate that in all cases SELF interventions are better than or equal
to OTHER interventions in curbing disease diffusion.  These conclusions remain
true for other values of the infectivity $\lambda>\lambda_c$ as we can
see from Supplementary Figures SF~1 to SF~4, where we present figures
analogous to Fig.~\ref{2dplot} and~\ref{2dplot2} for $\lambda = 1.0$ and
$\lambda=2.0$.  Interestingly, from Fig.~\ref{2dplot}(b) it turns out
that the difference between the two types of PI is maximal for very
small values of $\alpha$ (as expected) and in general, for intermediate,
$\alpha$-dependent values of the adoption fraction $f$.

The numerical solution of MF equations also allows us to determine the
height $I_m$ of the peak of incidence over time.
This is a crucial quantity that must be
kept as low as possible in order to avoid the saturation of health care
systems by an inflow of too many seriously ill individuals.
Fig.~\ref{2dplot}(c) shows clearly that also in this respect SELF PIs
perform better than the OTHER counterparts, as witnessed by the asymmetry
of the plot with respect to the diagonal.
The same conclusion is confirmed by Fig.~\ref{2dplot}(d):
$\ImOTHER-\ImSELF \ge 0$ for any value of $\alpha$ and $f$.
Adopting SELF interventions ``flattens the curve'' more effectively than the
adoption of OTHER PIs does.
A comparison between panels (b) and (d) in
Fig.~\ref{2dplot} reveals also that PI adoption affects differently
the total number of infected individuals with respect to the peak incidence.
In particular, the improved performance of SELF PIs is maximized for
$\alpha \to 0$ and $f\approx 0.3$ for what concerns the flattening of
the curve, while $f\approx 0.45$ for the total size of the outbreak.

Unanticipated effects can also be observed by comparing the performance of
OTHER PIs with a scenario with no interventions at all, and given by
$\ai = \aoo = 1$, see Fig.~\ref{2dplot2}. While it is expected that the
effect of PIs grows with efficacy (small $\alpha$) and
widespread PI adoption ($f$ close to 1) it is surprising to see that in a
large domain of the parameter space $\RNONE-\ROTHER$
is very close to zero (PIs do not significantly change the number of
people eventually infected) while $\ImNONE-\ImOTHER$
is rather large: OTHER PIs do not lead to an overall reduction of the
outbreak size but indeed substantially flatten the curve.

\begin{figure}[t]
  \centering
  \includegraphics[width=\columnwidth]{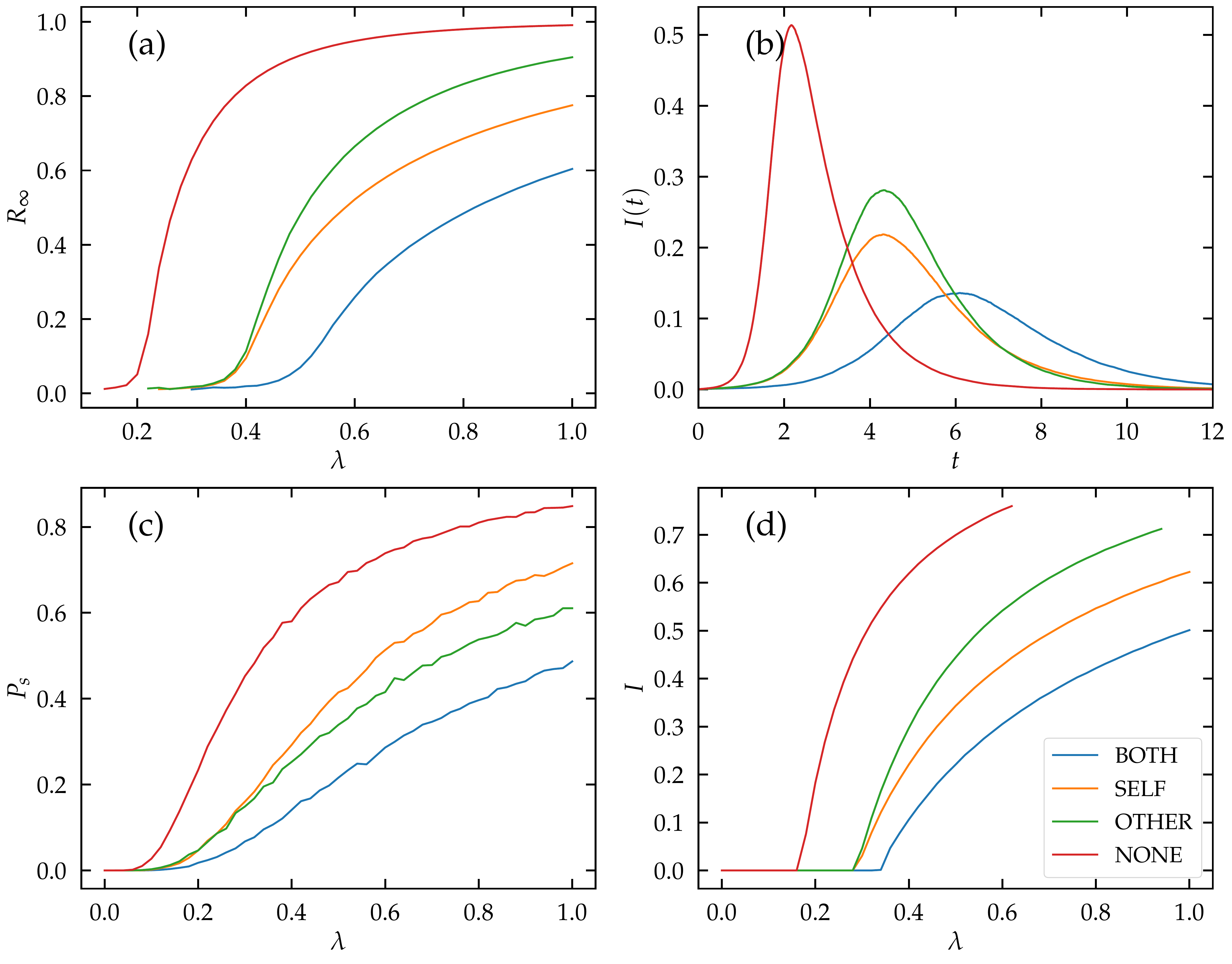}\\
  \caption{{\bf Numerical simulations confirm analytical results.}
    Results of numerical simulations on a random regular network of
    degree $K=7$.  The various curves are for the different scenarios
    of PI deployment.  In all cases $f=0.5$ and $\alpha=0.2$.  SIR
    dynamics: a) Value of the average size $R_\infty$ of extensive
    outbreaks as a function of $\lambda$; network size $N=10^4$,
    single random infected seed.  b) Temporal evolution of the
    incidence $I(t)$ for a single realization of the dynamics; network
    size $N=10^5$, 50 initial random infected seeds,
    $\lambda=0.5$. The exponential growth is the same for SELF and OTHER
    scenarios, but it lasts longer in the OTHER case.  c) Value of the
    probability $P_s$ to observe an extensive outbreak as a function
    of $\lambda$; network size $N=10^3$, single random infected seed.
    SIS dynamics: d) Value of the stationary density $I$ of infected
    nodes as a function of $\lambda$.  Network size $N=10^4$.}
  \label{fig:Rvslambda}
\end{figure}

The previous predictions have been obtained within the mean-field
framework.  The comparison with numerical simulations, to check whether
the picture derived above remains true, requires to take into account
the phenomenology of the stochastic SIR model in finite systems (see
Methods~\ref{FiniteN}), but reveals new interesting features.
Fig.~\ref{fig:Rvslambda}(a) shows that, also in numerical simulations,
the better efficacy of SELF interventions at the population level
remains valid for any value of the infectivity $\lambda$.  Indeed the
average size $R_\infty$ of extensive outbreaks is smaller when SELF PIs,
rather than OTHER PIs, are adopted.

Different values of the asymptotic prevalence for long times $R_\infty$ reflect
a different temporal evolution.  The mean-field approach predicts (see
Methods~\ref{sec:init-time-behav}) that the initial growth is
exponential, with a characteristic time scale depending only on the
threshold value, and thus on the product $\aoo \ai$, and hence is equal for both
SELF and OTHER scenarios with the same $\ai=\aoo=\alpha$.
The difference in the value of $R_\infty$ stems
from the fact that this exponential growth ends later in the OTHER case,
thus corresponding to a higher value of the peak incidence $I_{m}$.
Notice that this is different with what happens for the NONE ($\ai=1$,
$\aoo=1$) or BOTH ($\ai=\aoo<1$) scenarios.
In these other cases already the exponential growth rate is different.
Performing simulations of the stochastic SIR model (see Methods~\ref{FiniteN})
and analyzing the temporal evolution of the number of infected individuals
(see Fig.~\ref{fig:Rvslambda}(b)), these differences are clearly observed.
The maximum value of the
incidence in the SELF case is in general lower than in the OTHER case:
adopter-protecting interventions are more effective than contact-protecting
interventions in ``flattening the curve'' and thus reduce the maximum
instantaneous pressure on health care systems.

SELF interventions, however, are not the silver bullet against
epidemics. Indeed, while they reduce the overall prevalence of the
infectious disease, they are worse than OTHER interventions at
preventing the emergence of a macroscopic outbreak.  As
Fig.~\ref{fig:Rvslambda}(c) shows, the probability that a macroscopic
outbreak emerges is larger for the SELF prescription than for the OTHER
one.  This means that, if people adopt SELF PIs, it is more likely that
a single infected individual can give rise to a macroscopic outbreak.
However, as shown above, if such an outbreak occurs, or if the disease
is introduced in different nodes (for example because several infected
individuals arrive simultaneously in a community) the outbreak will
affect, on average, less people if SELF PIs are adopted.

\subsection{Recurrent diseases}

The paradigmatic model for recurrent diseases, leading to a steady
endemic state, is the Sus\-cep\-tible-Infected-Susceptible (SIS)
model~\cite{Keeling07book}.  In this case the epidemic transition
separates values of the parameter $\lambda=\beta/\mu$ for which the
disease quickly goes extinct ($\lambda \le \lambda_c$), from values such
that an endemic state is reached ($\lambda>\lambda_c$), with a steady
fraction of infected individuals.  Also in this case it is possible to
investigate the effect of different types of interventions via a
mean-field approach for homogeneous networks (see
Methods~\ref{hmf_sis}), obtaning for the threshold
\begin{equation}
  \label{eq:12}
  \lambda_c = \frac{1}{K} \frac{1}{f \ai \aoo + 1 - f},
\end{equation}
while the stationary density of infected nodes in the vicinity of this
threshold is
\begin{equation}
  I \simeq \frac{\Delta}{\lambda_c^2 K} \frac{f \ai + 1
    - f}{f \aoo \ai^2 +1 -f}.
\end{equation}
Hence the phenomenology of the SIS model is analogous to what is found
for SIR: The onset of the endemic state is a symmetric function of the
PI efficacies. The prevalence about it, however, is, for a constant $\aoo \ai$,
an increasing function of $\ai$ and a decreasing one of $\aoo$.
Interventions protecting the adopter (small $\ai$, $\aoo=1$)
result in lower disease circulation at the population level than
interventions protecting contacts of the adopter (small $\aoo$, $\ai=1$).
These predictions are verified
numerically to be true for all values of $\lambda>\lambda_c$, not only
close to the transition (see Fig.~\ref{fig:Rvslambda}(d)).

\subsection{An application to COVID-19 diffusion}

While we have considered so far extremely simple epidemic models on
idealized networks, the qualitative conclusions extend to more
complicated epidemic models and more general types of networks. For
example, the extension to heterogeneous networks with a general degree
distribution $P(k)$, presented in the Supplementary Information, leads to
results perfectly analogous to those discussed above. To give a more realistic
example, here we consider a Susceptible-Exposed-Infected-Recovered
(SEIR) dynamics on a large weighted network describing
interactions among people in Portland, OR~\cite{Hartnett2021}.
This network has been inferred using mobile device data before social
distancing measures were enacted during the COVID-19 pandemic.  SEIR
model parameters were calibrated to describe the first wave of COVID-19
in 2020 (see Methods~\ref{Portland} for more details).
The size of the
network is $N=214,393$, which is a substantial fraction of the census
population of Portland city in 2020, namely $N_P = 652,503$~\cite{population}.

\begin{figure}[t]
  \centering
  \includegraphics[width=\columnwidth]{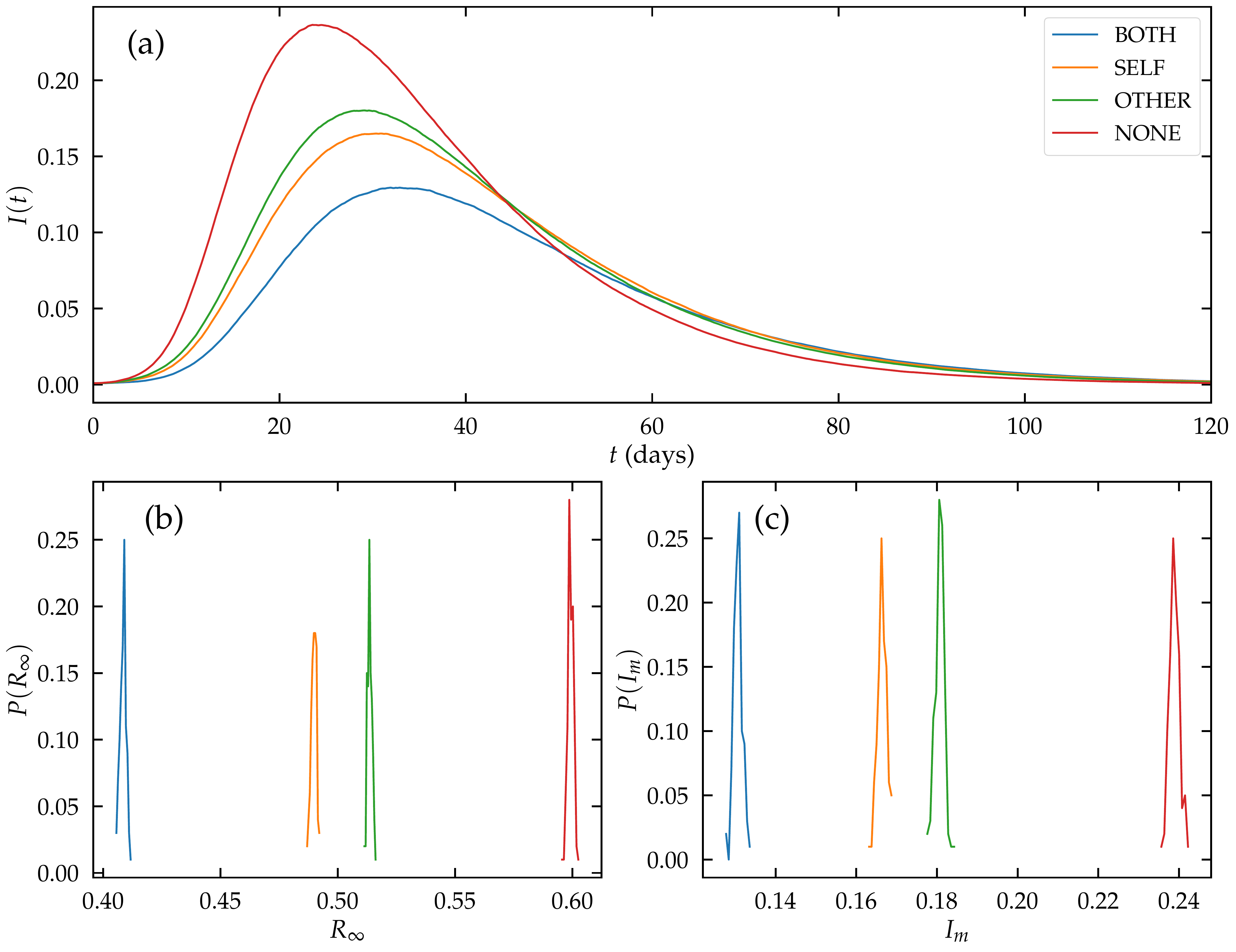}

  \caption{{\bf SEIR dynamics on the Portland network.}  a) $I$ vs $t$
    for a single run; b) Distribution of the fraction $R$ of recovered
    people in the asymptotic state for 100 runs in each of the 4
    scenarios; c) Distribution of the maximum fraction $I_m$ of
    simultaneously infected people for 100 runs in each of the 4
    scenarios.}
  \label{fig:Portland}
\end{figure}

In this framework, we tested the four scenarios for PI efficacy,
assuming again $f=0.5$.  Figure~\ref{fig:Portland} shows that also in
this case SELF interventions flatten the curve of the number of infected people
more effectively than OTHER interventions with the same individual efficacy.
With respect to the final prevalence $R_\infty$, adopting SELF PI reduces the
value in the absence of interventions $\RNONE = 0.595$ to $\RSELF=0.487$,
which is significantly (beyond stochastic fluctuations) smaller than
$\ROTHER=0.510$, while for interventions with both SELF and OTHER efficacy the
prevalence is $\RBOTH=0.406$.  On the scale of the city, extrapolating
these values for a population of $N_P = 652,503$, the adoption of SELF PI
instead of OTHER PI would imply a reduction of around $15,000$ in the
total number of cases. For the maximum number of simultaneously infected
people, we obtain the values $\ImNONE = 0.238$, $\ImSELF = 0.166$,
$\ImOTHER = 0.180$, and $\ImBOTH = 0.130$, which on the city scale implies
a reduction of around $9,000$ concurrently infected individuals when
SELF interventions are used instead of OTHER interventions.

Although we do not claim these numbers to be directly applicable to a
real world scenario, the results clearly show that the collective
effects leading to a better performance of SELF interventions with respect to OTHER
interventions at the population level are at work also for more complicated
epidemic models on realistic networks.

\section{Discussion}
\label{sec:discussion}

A vast range of protecting interventions of different type, both
pharmaceutical and non-phar\-ma\-ceutical, can be used to combat the
diffusion of an infectious disease.  Each intervention may in general
affect differently the infectivity and the susceptibility of the single
individual.

Here we have considered, in a
simple formalism based on disease propagation on networks, the effects
of generic PIs on the reduction of the total number of
infected individuals and of the maximum number of individuals
concurrently infected, the so-called goal of ``flattening the curve'',
to prevent the saturation of health care systems.
In the simple
setting adopted, PIs are characterized by two parameters which,
in the case of masks, gauge the level of viral particle filtration
offered from the mask wearer to the environment ($\aoo$), and in the
opposite direction ($\ai$).

The results presented above demonstrate that collective effects have surprising
consequences for what concerns the level of global protection guaranteed by
different kinds of intervention.  In agreement with many other publications, we
find, of course, that no matter how imperfect, the adoption of PIs reduces the
circulation of an infectious disease with respect to the case of no adoption,
and that a higher efficiency at the individual level guarantees a higher
protection for the whole community.  Surprisingly, however, we find that
interventions protecting the adopter (SELF PIs) or interventions protecting
mostly contacts of the adopter (OTHER PIs) have {\em not} the same effect, even
for equal efficacy of the individual intervention.  SELF PIs turn out to reduce
more the overall incidence of the infection and to ``flatten the curve'' more
efficiently, i.e., they reduce more drastically the peak value of the number of
individuals simultaneously infected.  Although we have focused on purely SELF
($\aoo=1$, $\ai<1$) and purely OTHER ($\aoo<1$, $\ai=1$) interventions, the same
picture applies when SELF (OTHER) PIs provide some small level of protection
also in the other direction, i.e.  $\ai \ll \aoo$ ($\ai \gg \aoo$), respectively.
Our results are presented in terms of a mean-field theory for the simple SIR and
SIS models of disease propagation in homogeneous networks, that can be readily
extended to heterogeneous contact patterns (see Supplementary Information), and
are backed up with numerical simulations. Additionally, we consider a SEIR
model, often used for COVID-19 analyses and estimations, on an empirical network
representing contact patterns in the city of Portland. This last analysis allows
us to confirm the expected effect of the different kinds of PIs in a realistic
scenario.  Our results can be interpreted in the framework of the general theory
for heterogeneous agents put forward by Miller~\cite{Miller2007}, predicting
that control strategies having a heterogeneous impact on susceptibilities are
more effective in reducing epidemic size while strategies with heterogeneous
impact on infectivities reduce more the probability of large outbreaks.

What causes the difference between the performance of the SELF and OTHER
strategies? To investigate this issue we considered a system where the choice of
the PI is not individual-based, but contact-based: in other words, $\beta_{ij}$
is still a quenched random variable given
by~\eqref{eq:def_infection_rate_directed}, but the variables $m_i$ and $m_j$ are
extracted independently for each edge.  Hence, considering the example of mask
wearing, an individual $i$ can use a mask when interacting with contact $j_1$
and not use it when interacting with contact $j_2$.  In such a case (see
Supplementary Figure SF5) SELF and OTHER scenarios give exactly the same
results.  Therefore asymmetry arises at the community level only if each
individual consistently adopts (or does not adopt) the protecting intervention
in every interaction.

Concerning the example of masks, our results place a strong question mark on
policies adopted to curb the diffusion of COVID-19.  During the initial stages
of the pandemic, concerns were raised about the use of valved masks, as it was
argued that they do not protect contacts~\cite{doi:10.1063/5.0031996}.  This led
to advice against their use by the World Health Organization
(WHO)~\cite{who_masks} and the Centers for Disease Control and Prevention
(CDC)~\cite{cdc_masks}, and to outright bans in some cities~\cite{valved-cities}
and in major US airlines~\cite{valved-airplanes}.  At the same time the usage of
masks protecting contacts of the wearer has been encouraged.  As we have shown,
unless effective masks providing protection in both directions can be widely
used, it is better to adopt SELF (valved) masks to maximally reduce the impact
of the disease at the population level.

This global effect is not the only argument pushing to reconsider this type of
policies. Another compelling reason has to do with the utility of single
individuals.  The usage of PIs can be imposed by ordinances issued by the
government, but the adoption of individual PIs, even if policed by authorities,
always involves a personal choice.  As shown above, the adoption of an
intervention that protects only the contacts of the adopter ($\ai=1$) does not
imply any direct utility for the adopter: The probability to be eventually
infected is exactly the same for adopting and nonadopting persons.  On the other
hand, adopting an intervention (for example wearing a mask) is a burden, so that
the net direct utility for adopting an OTHER PI is negative.  A positive overall
payoff for the individual adopting an OTHER PI can be obtained only if many
people cooperate so that the common good is achieved, overcoming the
inconvenience implied by the adoption.  Thus OTHER PIs are both less effective
at the population level and less likely to be adopted at the individual one.  No
such barrier to adoption exists for self-protecting (SELF) PIs: In such a case
it is not only beneficial at the collective level, but also in the selfish
interest of each individual to adopt a protecting intervention.  Hence, we
conjecture that the overall higher effectiveness of self-protecting
interventions at the community level will be further enhanced by the fact that
they are more likely to be spontaneously adopted.

\section{Methods}

\subsection{Mean-field theory for the SIR model}
\label{hmf_sir}

In the case of a homogeneous network, with all nodes sharing the same
degree $k = K$, a mean-field approximation considers that all nodes with
the same PI state are equivalent, and therefore we can characterize the
dynamics by the probability that a node in a given PI state (adopter,
nonadopter) is in the compartment $S$, $I$, or $R$.  Under these
conditions, the dynamics of the network is defined by the set of
probabilities
\begin{eqnarray}
  \label{eq:1}
  I_a, \quad S_a, \quad R_a,
\end{eqnarray}
that an individual in the PI state $a$, taking values $a=1$ (adopter)
and $a = 0$ (nonadopter), is infected, susceptible or recovered,
respectively.  In a population of fixed size, these sets of
probabilities are subject to the normalization condition
\begin{equation}
  I_a +  S_a +  R_a = 1.
  \label{eq:36}
\end{equation}

The mean-field equations for the system are constructed by considering the
contacts that can change the number of infected individuals in each
PI state. Thus, the rate equation for the number of removed
individuals takes the form
\begin{equation}
  \label{eq:4}
  \frac{d R_a}{d t} = \mu I_a,
\end{equation}
for all PI states. For the density of infected adopting individuals, we
can write~\cite{Keeling07book,PastorSatorras2015}
\begin{equation}
  \frac{d I_1}{d t} =-\mu I_1 + (K-1)  S_1
  \left[  f
    I_1 \beta \ai \aoo + (1-f) I_0  \beta
    \ai\right ].
  \label{eq:eq1}
\end{equation}
In this equation, the first term considers the decay to the recovered
state with rate $\mu$. The second term considers the probability of a
new infection along a given edge, that is proportional to the density
of susceptible adopting individuals, and the probability that
the edge points to an adopting infected individual (with probability $f$),
which will lead to an effective infection rate $\beta \ai \aoo$, or to an
infected nonadopting individual (with probability $1-f$), associated to an
effective infection rate $\beta \ai$. The term $K-1$ takes into
account that infected individuals have at most only $K-1$ edges
available to propagate the infection, since one is used up by the
neighbor that induced their infection. Rescaling time by $\mu$, we can
write
\begin{equation}
  \frac{d I_1}{d t} = - I_1 + \lambda \ai (K-1) S_1 \theta,
  \label{eveqin}
\end{equation}
where $\lambda = \beta/\mu$ and we have defined
\begin{equation}
  \theta =f  \aoo I_1  + (1-f) I_0.
  \label{eq:24}
\end{equation}
From here, the rate equation for adopting susceptible individuals is
\begin{equation}
  \label{eq:2}
  \frac{d S_1}{d t} = - \lambda \ai (K-1) S_1 \theta.
\end{equation}
Finally, for nonadopting individuals, we can write
\begin{eqnarray}
  \frac{d I_0}{d t} &=& -I_0 + \lambda (K-1) S_0 \theta, \\
  \frac{d S_0}{d t} &=& -\lambda (K-1) S_0 \theta.
                        \label{eveqfin}
\end{eqnarray}

From the general Eq.~(\ref{eq:4}), considering that $\mu$ is absorbed
into the time rescaling, we have
\begin{equation}
  R_a(t) = \int_0^t I_a(t') dt',
\end{equation}
and from Eqs.~(\ref{eq:2}) and~(\ref{eveqfin}),
\begin{equation}
  S_1(t) =  e^{- \lambda (K-1) \ai \phi(t)}, \quad S_0(t) = e^{
    - \lambda (K-1) \phi(t)},
  \label{eq:38}
\end{equation}
where $\phi(t) = \int_0^t \theta(t') dt'$ can be written as
\begin{equation}
  \phi(t) =  f \aoo R_1(t)  + (1-f) R_0(t),
\end{equation}
and where we assume an initial condition given by a vanishing fraction of
infected individuals.

\subsubsection{Threshold evaluation}
\label{threshold}

To estimate the value of the threshold, we consider the final outbreak
size, given by the number of removed individuals at time
$t \to \infty$.
Since $I_a(\infty) = 0$, we have from the
normalization condition, Eq.~(\ref{eq:36}),
$R_a(\infty) = 1 - S_a(\infty)$. In this infinite time limit,
defining $\phi_\infty \equiv \phi(\infty)$,
\begin{eqnarray}
  \label{eq:37}
  \phi_\infty &=& f \aoo R_1(\infty)  + (1-f) R_0(\infty) \\
              &=& f \aoo [1 - e^{- \lambda (K-1) \ai \phi_\infty}] \\
              &+& (1-f) [1 - e^{- \lambda (K-1)  \phi_\infty}]
                  \equiv \Psi(\phi_\infty).
\end{eqnarray}
A non-zero solution is obtained when
$\left .\frac{d \Psi(\phi_\infty)}{ d \phi_\infty}\right|_{\phi_\infty  = 0} \geq 1$,
  leading to the threshold condition
$\lambda > \lambda_c$ to observe a finite outbreak, with
\begin{equation}
  \label{eq:10}
  \lambda_c = \frac{1}{(K-1) \left[  f \ai \aoo +1 - f \right]}.
\end{equation}

\subsubsection{Behavior close to the threshold}
\label{PrevalenceSIR}

In the limit $\lambda \to \lambda_c^+$, we have that $R_a(\infty)$ and
$\phi_\infty$ are both small. From Eq.~(\ref{eq:37}), performing an
expansion for small $\phi_\infty$ up to second order, we can solve it
and obtain
\begin{equation}
  \label{eq:3}
  \phi_\infty \simeq \frac{2 \Delta}{\lambda_c^3 (K-1)^2}
  \frac{1}{f \aoo \ai^2 +1 - f},
\end{equation}
where $\Delta  = \lambda - \lambda_c$ is the distance to the
critical threshold.  For the total prevalence,
$R_\infty = f R_1(\infty) + (1-f)R_0(\infty)$, using Eq.~(\ref{eq:38})
and the normalization condition, we can write
\begin{eqnarray}
  \label{eq:39}
  R_\infty &=& f  [1 - e^{- \lambda (K-1) \ai \phi_\infty}] + (1-f) [1 -
               e^{- \lambda (K-1)  \phi_\infty}] \nonumber\\
           &\simeq& \lambda(K-1) [f \ai +1 - f] \phi_\infty,
\end{eqnarray}
where the have expanded the exponential factors to leading order.
Combining Eqs.~(\ref{eq:39}) and (\ref{eq:3}), we finally obtain
the total prevalence close to the threshold
\begin{equation}
  \label{eq:7}
  R_\infty \simeq  \frac{2 \Delta}{\lambda_c^2 (K-1)}\frac{f \ai +1 -
    f}{f \ai^2 \aoo + 1 - f}.
\end{equation}

\subsubsection{Relation between $R_0(\infty)$ and $R_1(\infty)$}
\label{relation}
Taking the limit $t \to \infty$ in~\eqref{eq:38} and eliminating
$\phi_\infty$ we obtain $S_1 = S_0^{\ai}$, from which, using the condition
$R_a(\infty) = 1 - S_a(\infty)$, we arrive at
\be
R_1(\infty) = 1 - [1-R_0(\infty)]^{\ai}.
\ee

\subsubsection{Initial time behavior}
\label{sec:init-time-behav}

To estimate the initial time behavior of the epidemic outbreak, we
consider the limit of a very small density of infected individuals.
Linearizing the corresponding equations, we have
\begin{eqnarray}
  \label{eq:8}
  \dot{I}_1 &\simeq& - I_1  +  \lambda (K-1) \ai \theta, \\
  \dot{I}_0 &\simeq& - I_0  +  \lambda (K-1)  \theta.
                     \label{eq:18}
\end{eqnarray}
Using these expressions in Eq.~(\ref{eq:24}), we have
\begin{equation}
  \label{eq:15}
  \dot{\theta} = f \aoo \dot{I}_1 + (1-f)\dot{I}_0 \simeq  -\theta
  + \frac{\lambda}{\lambda_c}\theta.
\end{equation}
Defining the characteristic time scale
\begin{equation}
  \label{eq:16}
  \tau = \frac{\lambda_c}{\lambda - \lambda_c},
\end{equation}
the solution for $\theta$ is
\begin{equation}
  \label{eq:17}
  \theta(t) = \theta(0) e^{t/\tau}.
\end{equation}
Using these expressions, we can integrate Eqs.~(\ref{eq:8})
and~(\ref{eq:18}) to obtain
\begin{eqnarray}
  \label{eq:19}
  I_1(t) &\simeq& \lambda_c(K-1) \ai \theta(0) e^{t/\tau} \\
  I_0(t) &\simeq& \lambda_c(K-1)  \theta(0) e^{t/\tau},
\end{eqnarray}
while the total incidence $I(t) = f I_1(t) + (1-f) I_0(t)$
has the form
\begin{equation}
  \label{eq:20}
  I(t) \simeq \lambda_c (K-1) \left[ f \ai + 1 - f \right]\theta(0)
  e^{t/\tau}.
\end{equation}
For the initial condition, let us consider that a randomly chosen node
is initially infected. In this case, $I_m(0) = I_0(0) = 1/N$,
and therefore
\begin{equation}
  \label{eq:21}
  \theta(0) = \frac{1}{N} \left[ f \aoo + 1 -f \right].
\end{equation}
Therefore
\begin{equation}
  \label{eq:22}
  I(t) \simeq \frac{\lambda_c (K-1)}{N} \left[ f \ai + 1 - f \right]  \left[
    f \aoo + 1 -f \right] e^{t/\tau}.
\end{equation}

\subsection{Phenomenology of the SIR model in finite networks}
\label{FiniteN}

\subsubsection{Two types of outbreaks}

Starting from a single infected node, outbreaks can develop in two
qualitatively different ways~\cite{kissbook}. Some of them die after a
few infection events; others instead survive much longer and affect an
extensive fraction of the individuals. This is reflected in the
distribution of outbreak sizes, which has two components.  For small
values of $\lambda$ only the small component exists, made up by
short-lived small outbreaks and weakly depending on $N$.  For large
values of $\lambda$ also the second component exists, peaked around a
size proportional to $N$.  The epidemic threshold marks the birth of the
large, extensive component.  This distinction is exactly defined only in
the thermodynamic limit (indeed the threshold is properly defined only
in this limit), but it is operatively meaningful also for finite, but
large, size $N$.  The fraction of outbreaks belonging to the large
component is the probability $P_s$ that an extensive outbreak
emerges. It is zero below the threshold and grows with
$\lambda>\lambda_c$ above it.  The average size of extensive outbreaks
$R_\infty$ (i.e. the average value calculated only for the large
component) is another observable that is zero up to $\lambda_c$ and
grows with $\lambda$.  For epidemics on undirected networks the two
quantities coincide~\cite{kissbook}. On directed networks (such as the
effective networks over which the epidemic spreads if PIs are used)
they are in general different~\cite{Kenah2007,Miller2007}.

By construction, the mean-field approach on
networks~\cite{PastorSatorras2015,kissbook} deals only with extensive
outbreaks in an infinite size system. Hence the quantity $R_\infty$
computed above has to be compared with the average size of macroscopic
outbreaks, which must be defined, in simulations, as outbreaks larger
than a given fraction of the whole system. We take this fraction to be
$0.01$. Clearly only for $N \to \infty$ and sufficiently far from the
transition point, results are independent from the choice of this
fraction.

\subsubsection{Temporal evolution}

For simulations in finite systems starting with a single infected node,
one must take into account that the exponential growth is preceded by a
regime dominated by stochastic fluctuations, whose duration has quite large
variations depending on the realization of the process. Only when a
sufficiently large fluctuation in the number of infected nodes $I(t)$ is
generated, exponential growth is triggered.  Averaging at fixed time
the value of $I(t)$ over many such realizations leads to a spurious
bending as a function of time.  Moreover, the duration of the initial
stochastic regime (and hence the size of the spurious bending) is
different in the various scenarios, further complicating the
analysis. To overcome these difficulties, as it is customary, we start
from a larger number of initial infected nodes, namely $50$, so
that the initial regime dominated by fluctuations is strongly
suppressed.

\subsection{Mean-field theory for the SIS model}
\label{hmf_sis}

The mean-field equations for the SIS model in a homogeneous network of
degree $k_i = K$ depend only on the densities of infected adopting and
nonadopting individuals $I_a$, and take the
form~\cite{Keeling07book,PastorSatorras2015}
\begin{eqnarray}
  \label{eq:23}
  \dot{I_1} &=& - I_1 + (1 - I_1) \lambda K \ai \theta \\
  \dot{I_0} &=& - I_0 + (1 - I_0) \lambda K  \theta,
                \label{eq:5}
\end{eqnarray}
with
\begin{equation}
  \label{eq:27}
  \theta =   f \aoo I_1 +    (1-f)I_0.
\end{equation}

\subsubsection{Threshold evaluation}
\label{sec:threshold-evaluation}

We can compute the threshold $\lambda_c$ by performing a linear stability
analysis around the solution $I_a = 0$, corresponding to the healthy,
non-endemic state. The Jacobian matrix of Eqs.~(\ref{eq:23})
and~(\ref{eq:5}), evaluated at the origin, is
\begin{eqnarray}
  \label{eq:6} J = \begin{pmatrix} -1 + \lambda K f \ai \aoo & \lambda K
(1-f) \ai \\ \lambda K f \aoo & \lambda K (1-f)
  \end{pmatrix},
\end{eqnarray} whose associated eigenvalues are $\Lambda_1 = -1$ and
$\Lambda_2 = -1 + \lambda K (f \ai \aoo +1 - f)$. The healthy state
becomes unstable when the largest eigenvalue becomes positive, that
is, $-1 + \lambda K (f \ai \aoo +1 - f) > 0$. This defines the
threshold $\lambda_c$, given by
\begin{equation}
  \label{eq:29}
  \lambda_c = \frac{1}{K}\frac{1}{f \ai \aoo +1 - f},
\end{equation}
such that for $\lambda > \lambda_c$ there is a steady infected state
with a non-zero prevalence.

\subsubsection{Behavior close to the threshold}

For homogeneous networks of degree $K$, the steady state condition
$\dot{I}_a = 0$ translates into the relations
\begin{eqnarray}
  \label{eq:30}
  I_1 &=& \frac{\lambda K \ai \theta}{1 + \lambda K \ai \theta},\\
  I_0 &=& \frac{\lambda K  \theta}{1 + \lambda K \theta}.
\end{eqnarray}
Inserting these into the definition of $\theta$, Eq.~(\ref{eq:27}), we
obtain the self-consistent equation
\begin{equation}
  \label{eq:9}
  \theta = \frac{\lambda K f \ai \aoo \theta}{1 + \lambda K \ai \theta}
  + \frac{\lambda K (1-f) \theta}{1 + \lambda K \theta}.
\end{equation}
Eq.~(\ref{eq:9}) can be solved close to the threshold, in the limit of
small $\theta$, performing a Taylor expansion up to second order,
which leads to the solution
\begin{equation}
  \label{eq:11}
  \theta \simeq \frac{\Delta }{\lambda_c^3 K^2} \frac{1}{f \aoo \ai^2 +1 -f}.
\end{equation}

Using the lowest order approximation $I_1 \simeq \lambda K \ai \theta$
and $I_0 \simeq \lambda K \theta$, the final steady state prevalence
$I = f I_1 + (1-f) I_0$ takes the form, close to the threshold
\begin{equation}
  \label{eq:32}
  I \simeq \frac{\Delta }{\lambda_c^2 K} \frac{f \ai + 1
    - f}{f \aoo \ai^2 +1 -f},
\end{equation}
in close analogy to the SIR result in Eq.~(\ref{eq:7}).

\subsubsection{Initial time behavior}

It is easy to see that the initial time behavior of the SIS model is
described by the same equations as the SIR case, simply replacing
the factor $K-1$ by $K$. Therefore, the time initial time evolution of
an outbreak initiated by a randomly chosen node takes the form, as
Eq.~(\ref{eq:22}),
\begin{equation}
  \label{eq:33}
  I(t) \simeq \frac{\lambda_c K}{N} \left[ f \ai + 1 - f \right]  \left[
    f \aoo + 1 -f \right] e^{t/\tau}.
\end{equation}

\subsection{Details on the network and the epidemic model for COVID-19}
\label{Portland}

We consider as the contact pattern for the SEIR dynamics the static contact
network determined in Ref.~\cite{Hartnett2021} for the city of Portland,
Oregon, before social distancing measures were enacted. The network
has $N=214393$ nodes and $M=1538092$ edges, so that the average degree
is $\av{k}=14.4$. Further details on network statistics can be found in the
original publication. The network is undirected and weighted, with each
contact weighted according to its duration.

On top of this network we perform simulations of the SEIR epidemic model
in continuous time, using the Gillespie algorithm.
The model is characterized by three parameters. An exposed node E
spontaneously becomes infectious (I) at a rate $\alpha$, while the
transition between the infectious state E to the recovered state R
occurs spontaneously at rate $\mu$.
In the absence of PI, an infectious node $i$ transmits
the infection to a susceptible neighbor $j$ at a rate $\beta w_{ij}$ where
$w_{ij}$ is the weight associated to the network edge.
In the presence of PIs this quantity is further modified as
in Eq.~\eqref{eq:def_infection_rate_directed}. The values of the parameters
are the same calibrated for COVID-19 in Ref.~\cite{Hartnett2021}:
$\alpha=1/3$, $\nu=1/14$, $\beta=1.337$. The initial condition for
simulations is that a fraction $10^{-3}$ of the individuals
is infected and the rest is susceptible.

\section*{Acknowledgments}

We acknowledge financial support from the Spanish
MCIN/AEI/10.13039/501100011033, under Project No. PID2019-106290GB-C21

\clearpage

\onecolumngrid
\renewcommand{\thefigure}{SM-\arabic{figure}}

\setcounter{figure}{0}
\section*{Supplementary Information}

\section{Heterogeneous mean-field theory for the SIR model}
\label{sec:heter-mean-field}

In the case of heterogeneous networks with a non-trivial degree
distribution, heterogeneous mean-field (HMF)
theory~\cite{PastorSatorras2015} considers that all nodes in the network
with the same degree share the same dynamical state, and therefore we
can characterize the state of the epidemics in terms of the probability
that a node of degree $k$ is in the state $S$, $I$, or $R$, taking the
same value for all nodes with the same degree. In the case of a
population with a fraction $f$ of protecting intervention (PI) adopters we
must consider the independent probabilities for adopting and nonadopting
individuals, in such
a way that the state of the network is fully described by the set of
probabilities
\begin{equation}
  I_k^1, \quad S_k^1, \quad R_k^1
\end{equation}
that an adopting individual of degree $k$ is infected, susceptible or recovered,
respectively, and the corresponding probabilities for nonadopting individuals
\begin{equation}
  I_k^0, \quad S_k^0, \quad R_k^0.
\end{equation}
Both sets of probabilities are subject to the normalization condition
\begin{equation}
   I_k^1 +  S_k^1 +  R_k^1 = I_k^0 + S_k^0 + R_k^0 = 1.
\end{equation}

In order to construct the rate equations for the time evolution of these
variables we need to consider the probability that a randomly chosen
link points to an infected node. From a mean field point of view,
considering that any individual has a probability $f$ to adopt a PI,
and assuming that the contact networks are undirected and
uncorrelated~\cite{alexei,Dorogovtsev2008} we can write the rate
equation for the density of infected adopting individuals of degree $k$
as~\cite{refId0,mariancutofss,PastorSatorras2015}
\begin{equation}
  \frac{d I_k^1}{d t} = -\mu I_k^1 + S_k^1 k
  \sum_{k'} \frac{(k'-1) P(k')}{\av{k}} \left[  f
    I_{k'}^1 \beta \ai \aoo + (1-f) I_{k'}^0 \beta
    \ai\right ].
  \label{eq:eq1}
\end{equation}
In this equation, the second factor considers the probability that a
link emanating from any node points to a node of degree $k'$ (taking
into account in the $-1$ term that an infected node is necessarily
connected to at least another infected node that transmitted the
disease to it and is thus not capable of further transmission), and the
probability that this node of degree $k'$ either adopts a PI (with
probability $f$) and the effective infection rate is $\beta \ai \aoo$, or
he/she does not adopt a PI (with probability $1-f$), with an effective
infection rate $\beta \ai$. Performing a time rescaling $t \to t \mu$,
Eq.~\eqref{eq:eq1} can be written as
\begin{equation}
  \frac{d I_k^1}{d t} = - I_k^1 + \lambda \ai S_k^1 k \theta,
  \label{eveqin}
\end{equation}
where $\lambda = \beta/\mu$ and we have defined
\begin{equation}
  \theta = f \aoo \sum_{k'} \frac{(k'-1) P(k')}{\av{k}} I_{k'}^1  +
  (1-f)  \sum_{k'} \frac{(k'-1) P(k')}{\av{k}} I_{k'}^0.
  \label{eq:24}
\end{equation}
From here, we can write the rate equations for the rest of variables
related to adopting individuals as
\begin{eqnarray}
  \frac{d S_k^1}{d t} &=& - \lambda \ai S_k^1 k \theta,\\
  \frac{d R_k^1}{d t} &=& I_k^1,
\end{eqnarray}
while the variables affecting nonadopting individuals fulfill the rate
equations
\begin{eqnarray}
  \frac{d I_k^0}{d t} &=& -I_k^0 + \lambda S_k^0 k \theta, \\
  \frac{d S_k^0}{d t} &=& - \lambda S_k^0 k \theta \\
  \frac{d R_k^0}{d t} &=& I_k^0.
  \label{eveqfin}
\end{eqnarray}
From these equations, we immediately have
\begin{equation}
  R_k^1(t) = \int_0^t I_k^1(t') dt', \quad  R_k^0(t) = \int_0^t I_k^0(t') dt'
\end{equation}
and
\begin{equation}
  S_k^1(t) =  e^{- \lambda k \ai \phi(t)}, \quad S_k^0(t) = e^{
  - \lambda k \phi(t)},
\end{equation}
where $\phi(t) = \int_0^t \theta(t') dt'$ can be written as
\begin{equation}
  \phi(t) = \sum_{k'} \frac{(k'-1) P(k')}{\av{k}}
  \left [  f \aoo R_{k'}^m(t)  + (1-f) R_{k'}^0(t)\right]
\end{equation}
and we assume that the initial condition consists in a very small
fraction of infected individuals and thus that almost all individuals are
susceptible.

In order to compute the epidemic threshold, we consider the state of the
system in the infinite time limit, where
$I_k^1(\infty) = I_k^0(\infty) = 0$. In this regime, the final
prevalence (density of infected individuals) is given by
\begin{eqnarray}
  R_k^1(\infty) &=& 1 - S_k^1(\infty) = 1 - e^{- \lambda k \ai
                    \phi_\infty} \label{eq:2}\\
  R_k^0(\infty) &=& 1 - S_k^0(\infty) = 1 - e^{- \lambda k
                    \phi_\infty} \label{eq:4},
\end{eqnarray}
with
\begin{eqnarray}
  \phi_\infty
  &=& \sum_{k'} \frac{(k'-1) P(k')}{\av{k}}
      \left [  f \aoo R_{k'}^1(\infty)  + (1-f) R_{k'}^0(\infty)\right] \nonumber\\
  &=& \frac{\av{k}-1}{\av{k}} [f \ai + 1 - f]
      - \sum_{k'} \frac{(k'-1)
      P(k')}{\av{k}} \left[ f \aoo e^{- \lambda k \ai \phi_\infty}  + (1-f)
      e^{- \lambda k  \phi_\infty}  \right]
      \equiv  \Psi(\phi_\infty).
\end{eqnarray}
A non-zero solution $\phi_\infty \neq 0$ is obtained when the condition
$\left .\frac{d \Psi(\phi_\infty)}{ d \phi_\infty}\right|_{\phi_\infty = 0} \geq 1$
is fulfilled, leading to
\begin{equation}
  \label{eq:1}
  \lambda  \frac{\av{k^2}-\av{k}}{\av{k}} \left[  f \ai \aoo +1 - f \right] \geq 1,
\end{equation}
which leads to the threshold value
\begin{equation}
  \label{eq:threshold}
  \lambda_c = \frac{\av{k}}{\av{k^2}-\av{k}} \frac{1}{ f \ai \aoo +1 - f}.
\end{equation}

From Eqs.~\eqref{eq:2} and~\eqref{eq:4} we can recover a relation
between $R_k^1(\infty)$ and $R_k^0(\infty)$, namely
\begin{eqnarray}
  \label{eq:3}
  R_k^1(\infty) = 1 - \left[ 1 -  R_k^0(\infty) \right]^{\ai},
\end{eqnarray}
analogous to the one obtained for homogeneous networks.

\section{Heterogeneous mean-field theory for the SIS model}
\label{sec:heter-mean-field-1}

The HMF equations for the SIS model~\cite{pv01a} in a general network
can be written in terms of the densities of infected adopting and
nonadopting individuals of degree $k$, $I_k^1$ and $I_k^0$, respectively,
that in this case take the form
\begin{eqnarray}
  \label{eq:23}
  \dot{I_k^1} &=& - I_k^1 + (1 - I_k^1) \lambda k \ai \theta \\
  \dot{I_k^0} &=& - I_k^0 + (1 - I_k^0) \lambda k  \theta,
\end{eqnarray}
where $\theta$, in random uncorrelated networks, is defined by
\begin{equation}
  \label{eq:27}
  \theta = \sum_{k'} \frac{k' P(k')}{\av{k}} \left[   f \aoo I_{k'}^1 +
    (1-f)I_{k'}^0 \right].
\end{equation}

The steady state condition, $\dot{I}_k^1 = \dot{I}_k^0 = 0$,
corresponding to the large time behavior, leads to the equations
\begin{equation}
  \label{eq:25}
  I_k^1 = \frac{\lambda \ai k \theta}{1 + \lambda \ai k \theta}, \qquad
  I_k^0 = \frac{\lambda  k \theta}{1 + \lambda  k \theta}.
\end{equation}
Inserting these expressions into the definition of $\theta$, we obtain the
self-consistent equation
\begin{equation}
  \label{eq:26}
  \theta = \sum_{k'} \frac{k' P(k')}{\av{k}} \left[ \frac{\lambda f \ai \aoo k'
      \theta}{1 + \lambda \ai k' \theta} + \frac{(1-f)\lambda k' \theta}{ 1 +
      \lambda k' \theta} \right] \equiv \Psi(\theta).
\end{equation}
The non-zero solution, corresponding to a finite prevalence, appears
when the condition
$\left. \frac{d \Psi(\theta)}{d \theta}\right|_{\theta=0} \geq 1$ is
fulfilled, which leads to
\begin{equation}
  \label{eq:28}
  \lambda \frac{\av{k^2}}{\av{k}} \left[ f \ai \aoo 1 - f \right] \geq 1,
\end{equation}
and to the threshold
\begin{equation}
  \label{eq:29}
  \lambda_c = \frac{\av{k}}{\av{k^2}} \frac{1}{f \ai \aoo + 1 - f}.
\end{equation}

\clearpage

\section{Supplementary figures}
\label{sec:suppl-figur}

\begin{figure}[h]
  \centering
  \includegraphics[width=0.98\columnwidth]{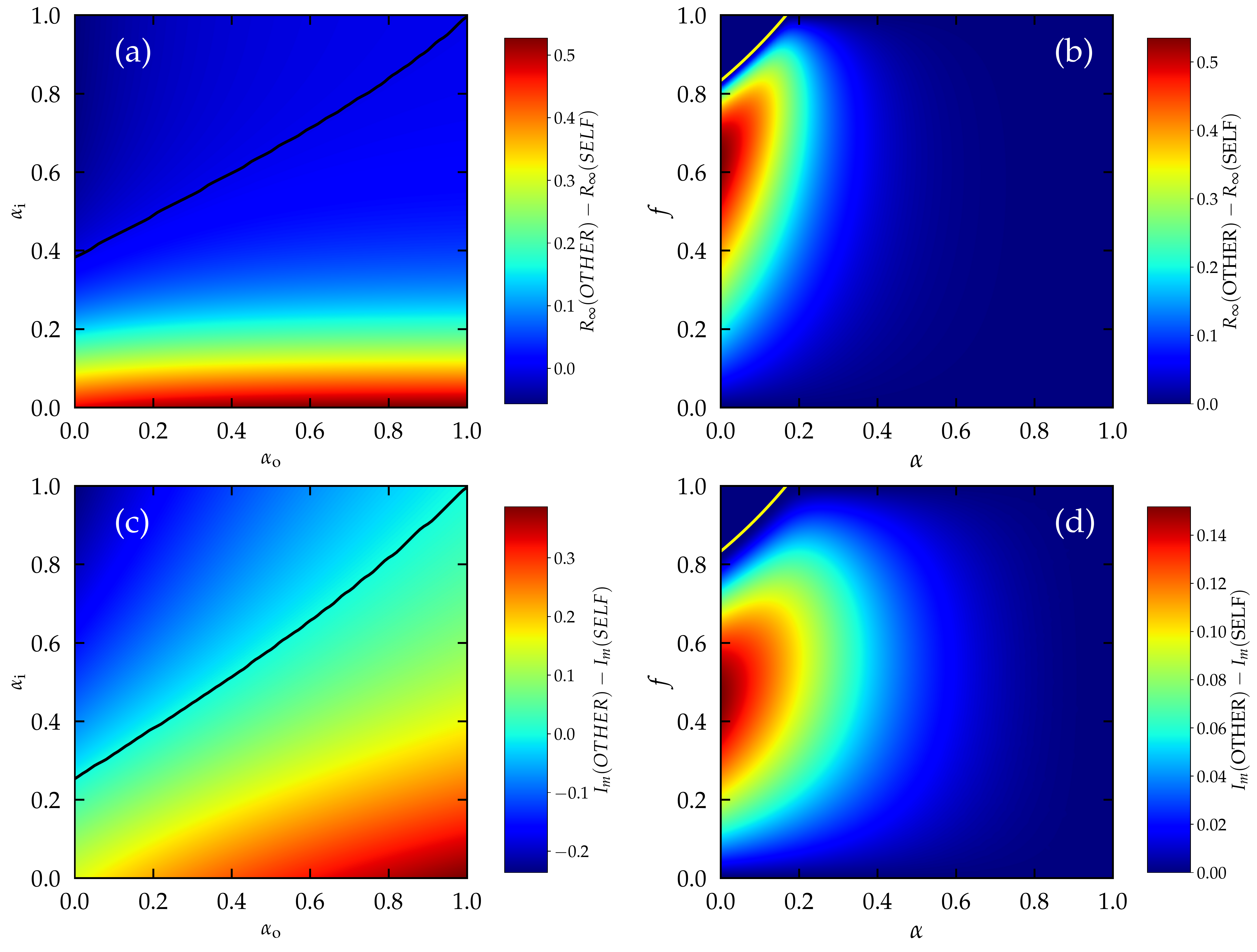}

  \caption{{\bf SELF interventions are more effective than OTHER interventions
      at the population level.}  (a) Difference between the final prevalence
    $R_\infty$ for the OTHER and the SELF scenario ($\ROUT-\RIN$), as a function
    of $\ai$ and $\ao$ of the individual PI. Notice that $\ROTHER$ depends only
    on $\ao$, and $\RSELF$ depends only on $\ai$..  The solid black line
    indicates the zero value.  (b) Difference between the final prevalence
    $R_\infty$ for the OTHER and the SELF scenario as a function of the fraction
    $f$ of adopters and the efficacy $\alpha$ of the individual PI.  The solid
    yellow line denotes where $\lambda_c(\alpha)=\lambda$.  Above it, the system
    is subcritical and $\ROUT-\RIN=0$.  (c) Difference between the maximum value
    of the incidence $I_{m}$ for the OTHER and the SELF scenario
    ($\ImOUT-\ImIN$), as a function of $\ai$ and $\ao$ of the individual
    PI. Notice that $\ImOTHER$ depends only on $\ao$, and $\ImSELF$ depends only
    on $\ai$.  The solid black line indicates the zero value.  (d) Difference
    between the maximum value of the incidence $I_{m}$ for the OTHER and the
    SELF scenario as a function of the fraction $f$ of adopters and the efficacy
    $\alpha$ of the individual PI.  The solid yellow line denotes where
    $\lambda_c(\alpha)=\lambda$.  In all plots quantities greater than zero
    indicate that SELF interventions perform better. Solution of homogeneous MF
    equations for a network of degree $K=7$ and for $\lambda=1.0$.  In panels
    (a) and (c) $f=0.5$. For these parameters, in panels (a) and (c) both SELF
    and OTHER strategies operate above their respective thresholds.}

\end{figure}

\begin{figure}[h]
  \centering
  \includegraphics[width=0.98\columnwidth]{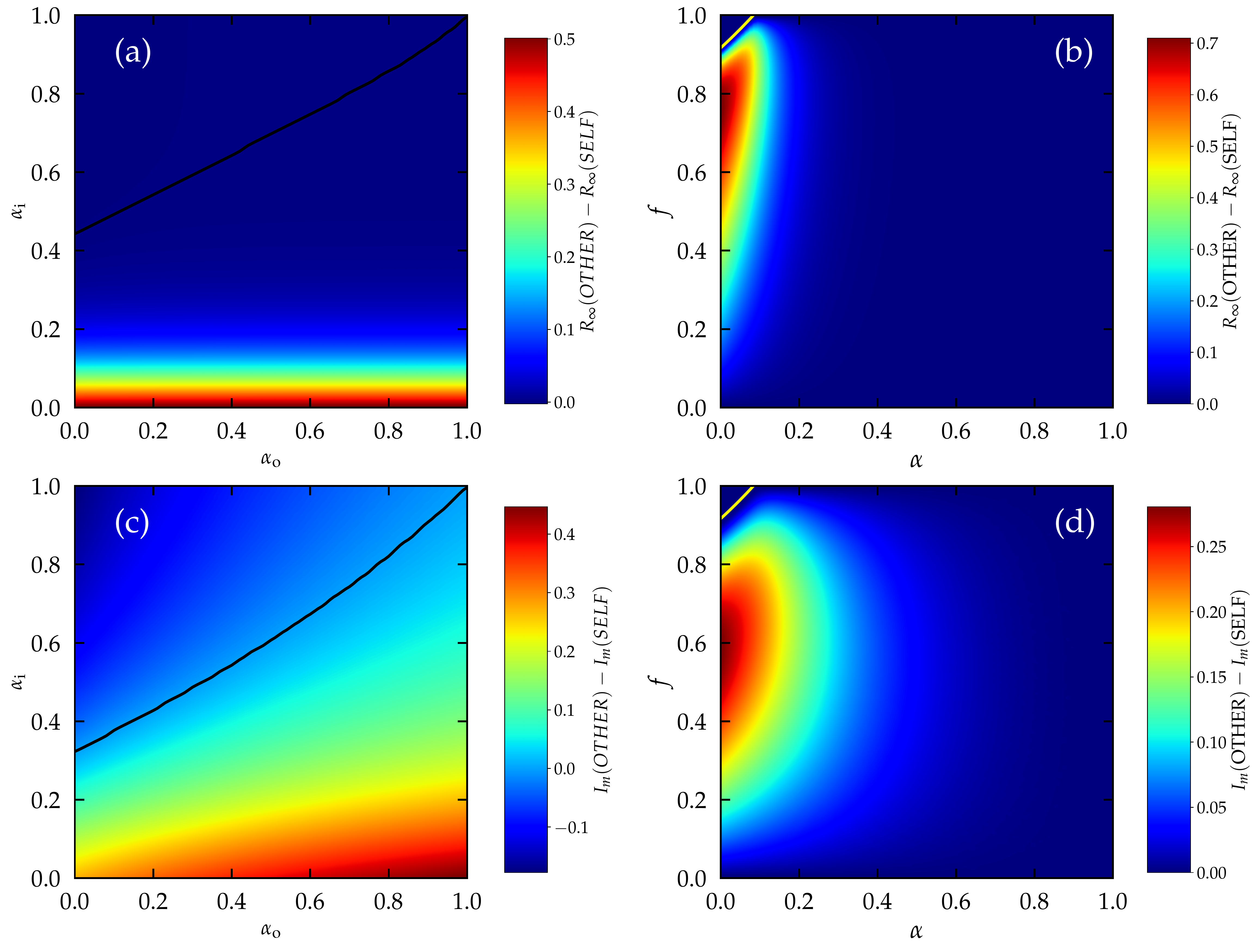}

  \caption{{\bf SELF interventions are more effective than OTHER interventions
      at the population level.}  (a) Difference between the final prevalence
    $R_\infty$ for the OTHER and the SELF scenario ($\ROUT-\RIN$), as a function
    of $\ai$ and $\ao$ of the individual PI.  Notice that $\ROTHER$ depends only
    on $\ao$, and $\RSELF$ depends only on $\ai$ The solid black line indicates
    the zero value.  (b) Difference between the final prevalence $R_\infty$ for
    the OTHER and the SELF scenario as a function of the fraction $f$ of
    adopters and the efficacy $\alpha$ of the individual PI.  The solid yellow
    line denotes where $\lambda_c(\alpha)=\lambda$.  Above it, the system is
    subcritical and $\ROUT-\RIN=0$.  (c) Difference between the maximum value of
    the incidence $I_{m}$ for the OTHER and the SELF scenario ($\ImOUT-\ImIN$),
    as a function of $\ai$ and $\ao$ of the individual PI. Notice that
    $\ImOTHER$ depends only on $\ao$, and $\ImSELF$ depends only on $\ai$.  The
    solid black line indicates the zero value.  (d) Difference between the
    maximum value of the incidence $I_{m}$ for the OTHER and the SELF scenario
    as a function of the fraction $f$ of adopters and the efficacy $\alpha$ of
    the individual PI.  The solid yellow line denotes where
    $\lambda_c(\alpha)=\lambda$.  In all plots quantities greater than zero
    indicate that SELF interventions perform better. Solution of homogeneous MF
    equations for a network of degree $K=7$ and for $\lambda=2$.  In panels (a)
    and (c) $f=0.5$. For these parameters, in panels (a) and (c) both SELF and
    OTHER strategies operate above their respective thresholds.}
\end{figure}

\begin{figure}
  \centering
  \includegraphics[width=0.98\columnwidth]{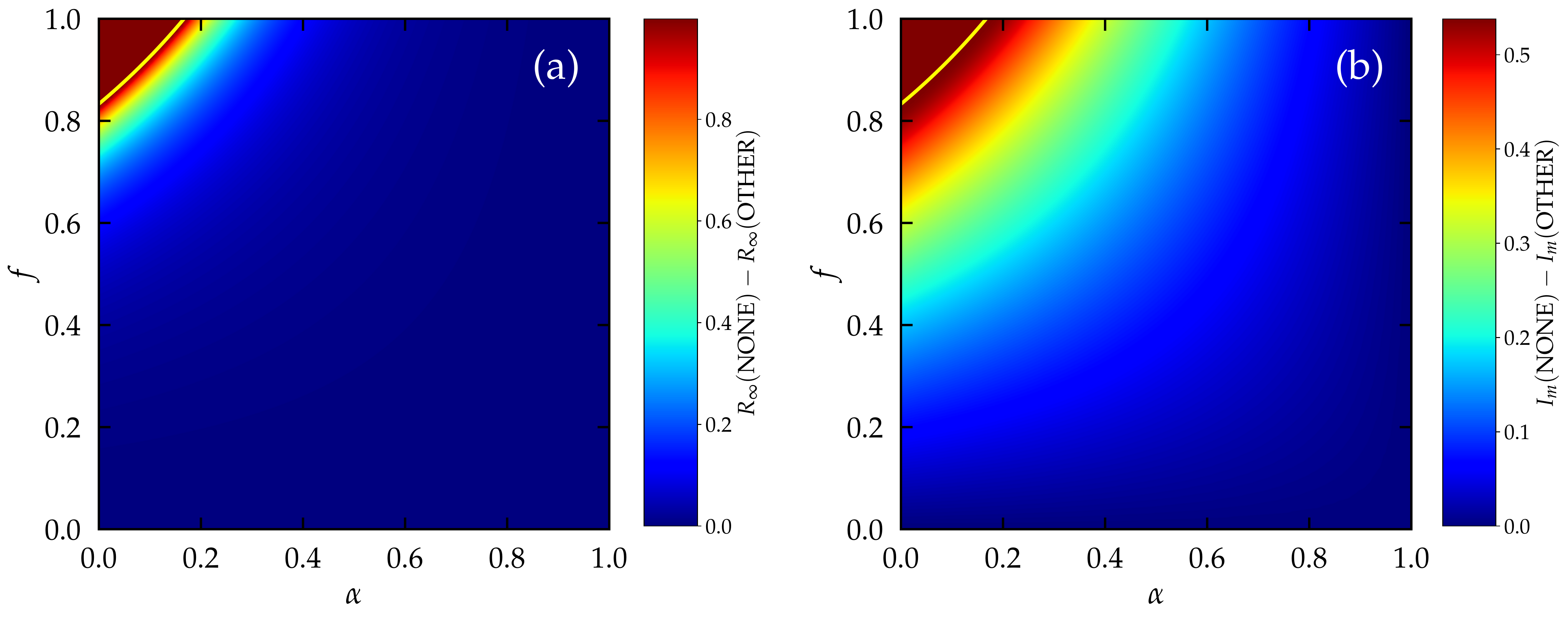}

  \caption{{\bf OTHER interventions flatten the curve but barely reduce
      prevalence with respect to no intervention.} (a) Difference
    between the final prevalence $R$ for the NONE and the OTHER scenario
    as a function of the fraction $f$ of adopters and the efficacy
    $\alpha$ of the individual PI.  (b) Difference between the maximum
    value of the incidence $I_{m}$ for the NONE and the OTHER scenario
    as a function of the fraction $f$ of adopters and the efficacy
    $\alpha$ of the individual PI. Solution of MF equations for a
    homogeneous network of degree $K=7$ for $\lambda=1$.  The solid
    yellow lines are where $\lambda_c(\alpha)=\lambda$ for the OTHER
    scenario.}

\end{figure}

\begin{figure}
  \centering
  \includegraphics[width=0.98\columnwidth]{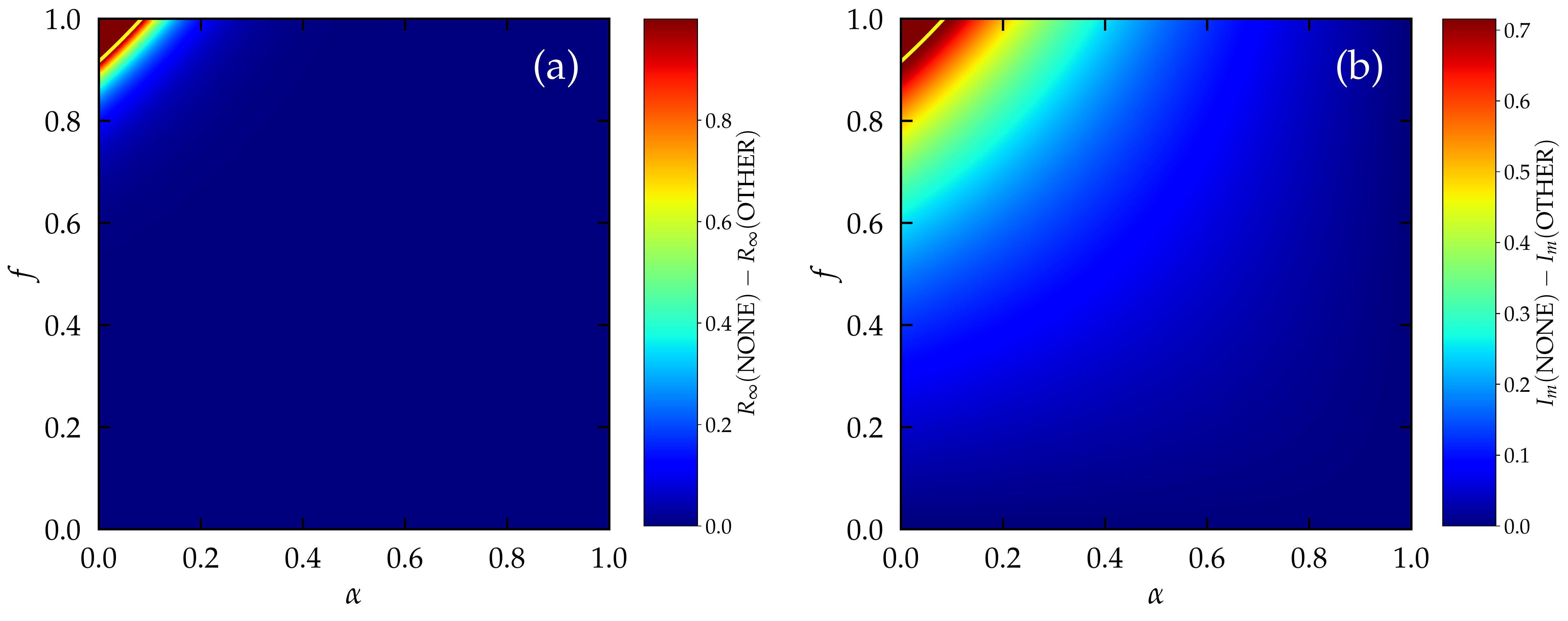}

  \caption{{\bf OTHER interventions flatten the curve but barely reduce
      prevalence with respect to no intervention.} (a) Difference
    between the final prevalence $R$ for the NONE and the OTHER scenario
    as a function of the fraction $f$ of adopters and the efficacy
    $\alpha$ of the individual PI.  (b) Difference between the maximum
    value of the incidence $I_{m}$ for the NONE and the OTHER scenario
    as a function of the fraction $f$ of adopters and the efficacy
    $\alpha$ of the individual PI. Solution of MF equations for a
    homogeneous network of degree $K=7$ for $\lambda=2$.  The solid
    yellow lines are where $\lambda_c(\alpha)=\lambda$ for the OTHER
    scenario.}

\end{figure}

\begin{figure}
  \centering
  \includegraphics[width=0.98\columnwidth]{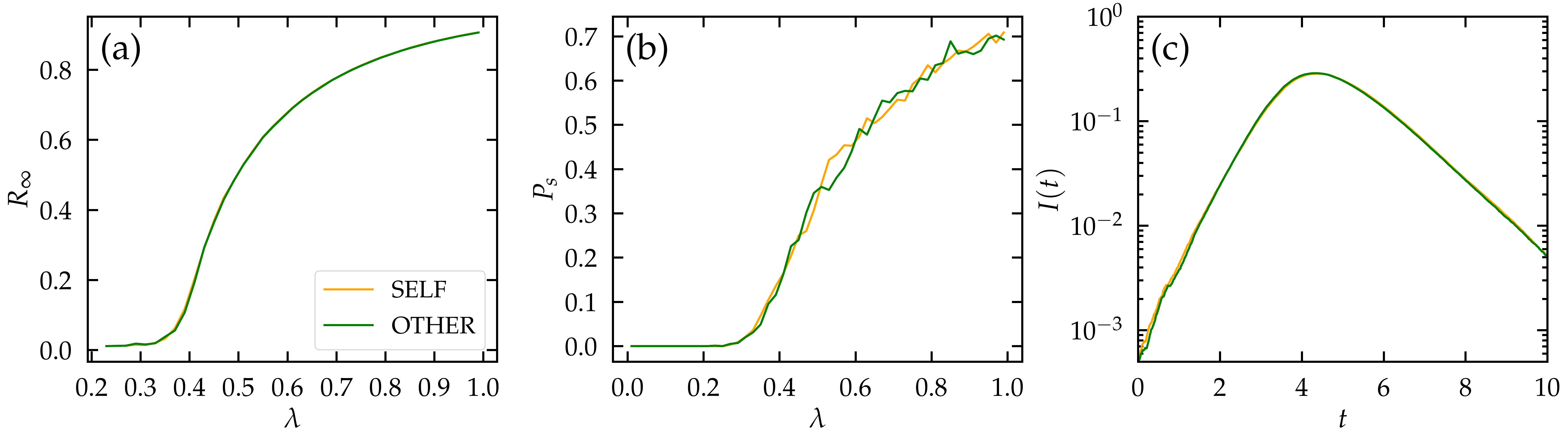}

  \caption{{\bf No difference between SELF and OTHER scenarios for
      contact-based protecting behavior.}  Simulations on a random
    regular graph with $K=7$, $N=10000$, $f=0.5$, $\alpha=0.2$. (a)
    Final prevalence $R_\infty$ of extensive outbreaks as a function of
    $\lambda$, for the SELF and the OTHER scenarios.  (b) Probability
    $P_s$ to observe an extensive outbreak as a function of $\lambda$
    for the SELF and the OTHER scenarios.  (c) Temporal evolution of the
    incidence $I(t)$ for a single realization of the dynamics;
    $N = 10^5$, 50 initial random infected seeds.  }
\end{figure}

\end{document}